\documentclass[fonts]{icst}

\usepackage{moreverb}
\usepackage[breaklinks,colorlinks,bookmarksopen,bookmarksnumbered,linkcolor=ICSTblue,citecolor=blue,urlcolor=ICSTblue]{hyperref}
\usepackage{breakurl}
\usepackage{doi}
\usepackage{amssymb}
\usepackage{amsmath}
\usepackage{bm}
\usepackage{graphicx}
\usepackage{array}
\usepackage{rotating}   
\usepackage{multirow}
\usepackage{makecell}
\usepackage{xcolor}
\usepackage{colortbl}
\usepackage{booktabs}
\usepackage{url}
\usepackage{paralist}
\setdefaultenum{\color{black} a)}{}{}{}

\definecolor{lightgray}{gray}{0.75}
\definecolor{lightred}{cmyk}{0, 0.206, 0.13, 0}
\definecolor{lightyellow}{cmyk}{0, 0, 0.1, 0}

\newcommand{\varDash}[1]{{\operatorname{\mathit{#1}}}}

\usepackage{subfig}
\newcommand{\specialcell}[2][c]{%
	\begin{tabular}[#1]{@{}l@{}}#2\end{tabular}}

\newcommand\BibTeX{{\rmfamily B\kern-.05em \textsc{i\kern-.025em b}\kern-.08em
T\kern-.1667em\lower.7ex\hbox{E}\kern-.125emX}}

\def\volumeyear{2014}

\journalname{XXXXXX}
\articletype{Research Article/Editorial}
 \def\volumeyear{XXXX}
 \def\volumenumber{XX}
  \def\issuemonth{XX-XX}
  \def\issuenumber{X}
  \def\articleid{XX}
  \def\copyrightholder{Author Name\textit{ et al.}, licensed to ICST}
  \copyrightnote{This is an open access article distributed under the terms of the Creative Commons Attribution license (\url{http://creativecommons.org/licenses/by/3.0/}), which permits unlimited use, distribution and reproduction in any medium so long as the original work is properly cited.}
  \def\articledoi{10.4108/XX.X.X.XX}
\received{XXXX}
  \accepted{XXXX}
  \published{XXXX}

\begin{document}

\runningheads{Ivan Homoliak et al.}{Improving Network Intrusion Detection Classifiers: An Adversarial Approach}

\title{Improving Network Intrusion Detection Classifiers by Non-payload-Based Exploit-Independent Obfuscations: An Adversarial Approach}

\author{Ivan Homoliak,\affil{1}$^,$\affil{2} Martin Tekn\"{o}s,\affil{1}  Mart\'{i}n Ochoa,\affil{3}$^,$\affil{4} Dominik Breitenbacher,\affil{1} Saeid Hosseini,\affil{2} Petr Hanacek\affil{1}}

\address{
	\affilnum{1}Faculty of Information Technology, Brno University of Technology, Bozetechova 1/2, 612 66 Brno, Czech Republic\\
	\affilnum{2}STE-SUTD Cyber Security Laboratory, 8 Somapah Road, 487372, Singapore\\
	\affilnum{3}Department of Applied Mathematics and Computer Science, Universidad del Rosario, Bogot\'{a}, Colombia\\
	\affilnum{4} Cyxtera Technologies
}

\abstract{
	Machine-learning based intrusion detection classifiers are able to detect unknown attacks, but at the same time they may be susceptible to evasion by obfuscation techniques. 
	An adversary intruder which possesses a crucial knowledge about a protection system can easily bypass the detection module.
	The main objective of our work is to improve the performance capabilities of intrusion detection classifiers against such adversaries. 
	To this end, we firstly propose several obfuscation techniques of remote attacks that are based on the modification of various properties of network connections; then we conduct a set of comprehensive experiments to evaluate the effectiveness of intrusion detection classifiers against obfuscated attacks. 		
	We instantiate our approach by means of a tool, based on NetEm and Metasploit, which implements our obfuscation operators on any TCP communication.
	This allows us to generate modified network traffic for machine learning experiments employing features for assessing network statistics and behavior of TCP connections.
	We perform evaluation on five classifiers: Gaussian Na{\"i}ve Bayes, Gaussian Na{\"i}ve Bayes with kernel density estimation, Logistic Regression, Decision Tree, and Support Vector Machines.
	Our experiments confirm the assumption that it is possible to evade the intrusion detection capability of all classifiers trained without prior knowledge about obfuscated attacks, causing an exacerbation of the TPR ranging from $7.8\%$ to $66.8\%$.
	Further, when widening the training knowledge of the classifiers by a subset of obfuscated attacks, we achieve a significant improvement of the TPR by $4.21\%$ -- $73.3\%$, while the FPR is deteriorated only slightly ($0.1\%$ -- $1.48\%$).
	Finally, we test the capability of an obfuscations-aware classifier to detect unknown obfuscated attacks, where we achieve over $90\%$ detection rate on average for most of the obfuscations.
}



\fnotetext[1]{Corresponding author.  Email: \email{ihomoliak@fit.vutbr.cz}}

\maketitle

	\section{Introduction}
Network intrusion attacks such as exploiting unpatched services continue to be one of the most dangerous threats in the domain of information security~\cite{mcafeeIntrusions},~\cite{certTargetVulns2015}.
Due to an increasing sophistication in the techniques used by attackers, misuse-based/knowledge-based~\cite{debar2000revised} intrusion detection suffers from undetected attacks such as zero-day attacks or polymorphism, enabling an exploit-code to avoid positive signature matching of the packet payload data. 
Therefore, researchers and developers are motivated to design new methods to detect various versions of the modified network attacks including the zero-day ones.
These goals motivate the popularity of Anomaly Detection Systems (ADS)
and also the classification approaches in the context of intrusion detection. 
Anomaly-based approaches are based on building profiles of normal users and trying to detect anomalies deviating from these profiles~\cite{debar2000revised}, which might lead to detection of unknown intrusions, but on the other hand it might also generate many false positives.
In contrast, the classification approaches take advantage of both misuse-based and anomaly-based models in order to leverage their respective advantages. 
The classification detection methods firstly build a model based on the labeled samples from both classes -- intrusions and the legitimate instances.
Secondly, they compare a new input to the model and select the more similar class as the predicted label.
Classification and anomaly-based approaches are capable to detect some unknown intrusions, but at the same time they may be susceptible to evasion by obfuscation techniques.

\vspace{-0.2cm}
\paragraph{Scope and Assumptions:}
Due to efficiency reasons as well as pervasive encryption, we assume in this work a classification-based network intrusion detection system that does not perform deep packet inspection and its model works with TCP connections objects, not single packets. 
Also, we assume an adversary who knows design details of such a system,\footnote{Note that when designing a protection system, this kind of adversary is stronger than adversary who does not know design details of a protection system (i.e., Kerckhoffs's principle).} but cannot modify its training data. 
The adversary can only modify the input of the system in a limited way that has to conform the protocol specification of the TCP/IP stack including victim's application.
The adversary can achieve it in several ways: a modification of exploit code, adding padding at the application layer of exploit code, artificially influencing network or transport layer protocols.
If an adversary wants to take advantage of huge database of existing exploits to make their obfuscated mutations and massively exploit targets, adding padding or various changes to exploit code, it may be time consuming and unsustainable with newly obtained exploits.\footnote{Note that the same holds for payload-based NIDS methods -- they need to spend a lot of effort to keep they database up-to-date with millions of modified attack samples daily.}
Therefore, the easiest way for an adversary is to design non-payload-based obfuscation techniques working at network and transport layers, which will mutate instances of known intrusions in an \textit{exploit-independent} way. 
This will make attacks similar to a legitimate traffic.
We follow this idea in our paper and construct exploit-independent modifications of attacks at network and transport layers of TCP/IP.
According to the taxonomy of adversarial attacks against IDS~\cite{corona2013adversarial}, our adversarial approach belongs to \textit{evasions of measurement phase} of IDS.
Considering \textit{influence}, \textit{security violation}, and \textit{specificity} as dimensions of taxonomy of attacks against learning systems~\cite{barreno2010security}, our obfuscated attacks belong to: 1) \textit{exploratory attacks}, which exploit misclassification but do not affect training data, 2) \textit{integrity attacks}, which compromise assets via false negatives, and 3) \textit{indiscriminate attacks}, which compromise wide class of instances.	

Despite the fact that non-payload-based evasions and obfuscations of network attacks are not  new research topics~\cite{handley2001network},~\cite{ptacek1998insertion},~\cite{puppy1999whiskers}, they are still challenging subjects -- to this date, this is witnessed by a few citations of the Stonesoft's technical report~\cite{boltz2010new}, which for the first time successfully applied non-payload-based obfuscations for existing network attacks.
There exist several related works considering non-payload-based adversarial evasions of network attacks for payload-based intrusion detection~\cite{ptacek1998insertion},~\cite{rubin2004automatic},~\cite{watson2004protocol}.
However, to the best of our knowledge, there are no studies on non-payload-based intrusion detection and obfuscation-based adversarial evasion (except our previous research~\cite{homoliak:NBAofObfNetVul,homoliak:ChofBOAinHTTP,2016-ihomoliak-thesis}).

\vspace{-0.2cm}
\paragraph{Problem statement:} In this work we address the following questions: 
\begin{compactitem}
	\item Is it possible to evade the detection of a non-payload-based intrusion detection classifier by obfuscation techniques?	
	
	\item If so, is it possible to increase the resilience of such a classifier against obfuscated attacks, or even detect unknown ones? 
\end{compactitem}

\vspace{-0.2cm}
\paragraph{Proposed Solution:} To address this problem, we define a set of obfuscation operators based on non-payload-based modifications of connection-oriented communications accomplished by \textit{NetEm} utility~\cite{hemminger2005netem} and \textit{ifconfig} command. 
Subsequently, we propose several experiments to train a classifier using obfuscated attacks as well as obfuscated legitimate connections and compare it against another model of a classifier that is unaware of obfuscated attacks.

\vspace{-0.2cm}
\paragraph{Contributions:} The main contributions of this paper are as
follows:
\begin{compactenum}
	\item We define non-payload-based obfuscation techniques and their influence on a classification task in an intrusion detection classifier.
	
	\item We implement several obfuscation techniques as part of our obfuscation tool and later conduct a data collection experiment that employs the obfuscation tool.
	\item We perform an evaluation of non-payload-based obfuscation techniques using our dataset, and we reveal them as: \textbf{1)} successful in evading detection by five classifiers that leverage selected subset of network connection features designed in~\cite{homoliakasnm2013}, as well as \textbf{2)} successful in an improvement of evasion resistance of the classifiers against unknown obfuscated attacks. 
	
	\item Moreover, we elucidate an alternative view on the outcome of our results, which is denoted as training data driven approximation of a network traffic normalizer.
	\item The collected dataset is provided to the research community.
\end{compactenum}

	\section{Background}\label{sec:MethodDescription}
Consider a session of a protocol at the application layer of the TCP/IP stack that serves for data transfer between the client/server based application.
The interpretation of such application data exchanges between client and server can be formulated, considering the TCP/IP stack up to the transport layer, by connection $k$ 
that is constrained to connection oriented protocol TCP at L4, Internet protocol IP at L3 and Ethernet protocol at L2.
The TCP connection $k$ is represented by start and end timestamps, ports of the client and the server, IP addresses of the client and the server, sets of packets sent by the client $P_{c},$ and by the server $P_{s}$, respectively.

\subsection{Features Extraction}
At this time, we can express characteristics of a TCP connection by network connection features. 
The features extraction process is defined as a function that maps a connection $k$ into space of features $F$: 
\begin{eqnarray}
     \begin{split}           
        f(k) &\mapsto F, \\
        F = (F_1, ~&F_2, \ldots, F_n), \\
     \end{split}
\end{eqnarray}
where $n$ represents the number of defined features.
Each function $f_i$ that extracts feature $i$ is defined as a mapping of a connection $k$ into feature space~$F_i$: 
\begin{eqnarray}
     f_i(k) \mapsto F_i, ~~i \in \{1,\ldots,n\},      
\end{eqnarray}
and each element\footnote{Representing a particular dimension of a feature.} of codomain $F_i$ is defined as 
\begin{eqnarray}
\label{featureElementDefinition}
  \begin{split}
    e = (e_0,\ldots,&e_n), ~n \in \mathbb{N}_0,\\
    e_i \in \mathbb{N} ~\mid ~e_i \in \mathbb{R} ~\mid e_i \in & ~\Gamma^{+}, ~i \in \{0,\ldots,n\},\\
    \Gamma = \{a-z, A&-Z, 0-9\},
  \end{split}
\end{eqnarray}
where $\Gamma^{+}$ denotes positive iteration of the set $\Gamma$.
In the context of this work, examples of such features are show in Table~\ref{FFS-features} of Appendix.

\subsection{Intrusion Detection Classification Task}\label{ADSPredictionTask}
Referring to~\cite{kohavi1995study}, let $X = V \times Y$ be the space of labeled samples,\footnote{A sample refers to the vector of the network features extracted over a connection.} where $V$~represents the space of unlabeled samples and $Y$ represents the space of possible labels.
Let $D_{tr} = \{x_1, x_2, \ldots, x_n\}$ be a training dataset consisting of $n$~labeled samples, where $x_i = (v_i \in V,~y_i \in Y)$.
Consider classifier $C$ which maps unlabeled sample $v \in V$ to a label $y \in Y$: 
\begin{eqnarray}
    y = C(v),
\end{eqnarray}
and learning algorithm $A$ which maps the given dataset $D$ to a classifier $C$: 
\begin{eqnarray}
C = A(D_{tr}).
\end{eqnarray}
The notation $y_{predict} = A(D_{tr}, v)$ denotes the label assigned to an unlabeled sample $v$ by the classifier $C$, build by learning algorithm $A$ on the dataset $D_{tr}$.
Now, all extracted features of the connection $k$ can be used as an input of the trained classifier $C$ that predicts the target label:
\begin{eqnarray}
y_{predict} = A\big(D_{tr}, f(k)\big),
\end{eqnarray}
where $y_{predict} \in \{\mathit{Intrusion}, \mathit{Legitimate}\}.$

\section{Proposed Approach}\label{ProposedApproach}
Considering the background from the previous section, now we describe non-payload-based obfuscations that aim at modification of the behavioral characteristics of a remote attack connection, and thus can influence the outcome of the intrusion detection classification task.

\subsection{Non-Payload-Based Obfuscations}
Consider connection $k_{a}$ representing a remote attack communication executed without any obfuscation.
Then, $k_{a}$ can be represented by features 
\begin{eqnarray}
f(k_a) \mapsto F^a = (F_1^a, F^a_2, \ldots, F^a_n),
\end{eqnarray}
which are delivered to the previously trained classifier $C$. 
Assume that $C$ can correctly predict the target label as an intrusive one, because its knowledge base is derived from training dataset $D_{tr}$ containing intrusive connections having similar (or the same) behavioral characteristics.

\begin{figure*}[!t]
	\centering{}    
	\includegraphics[width=11cm]{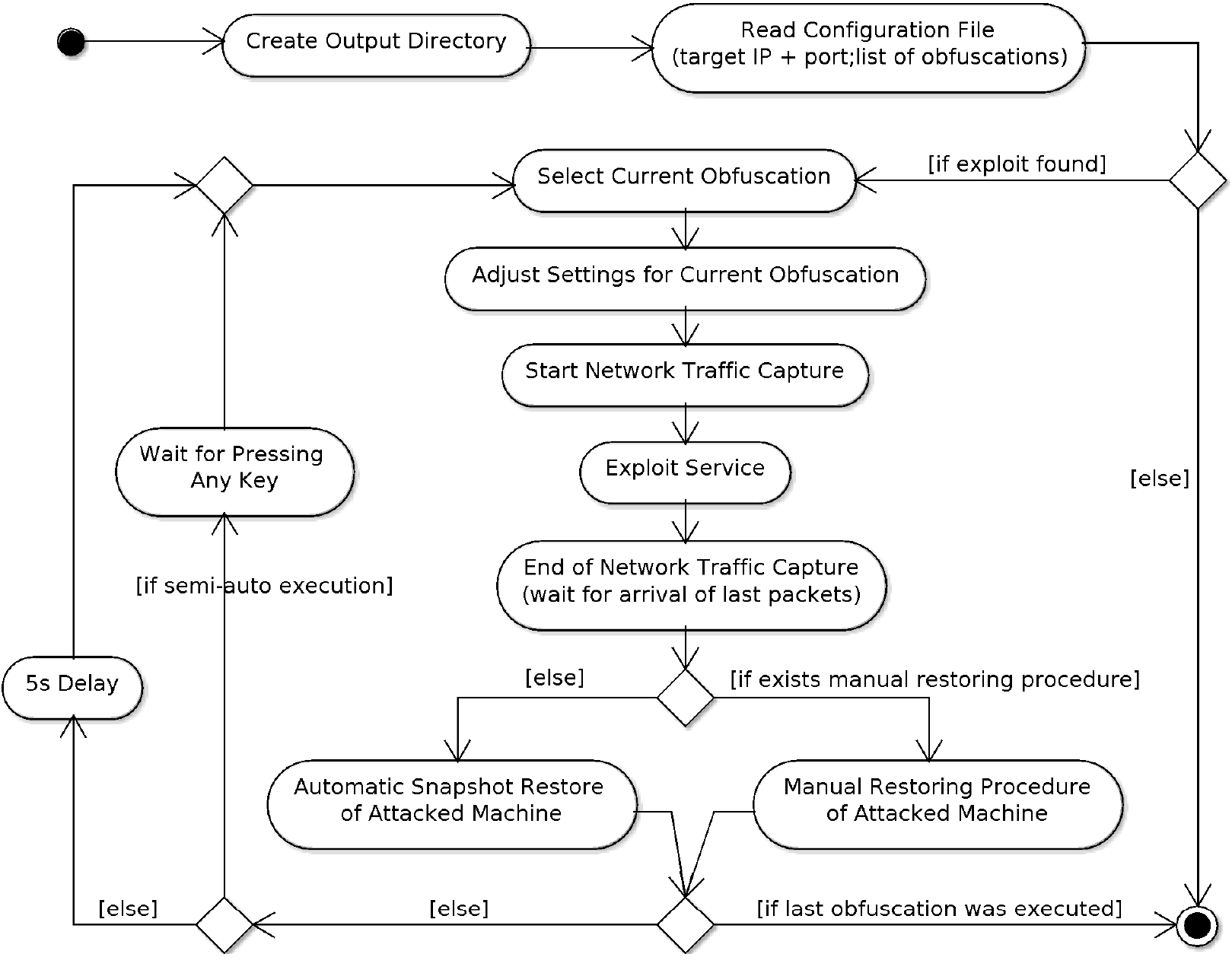}
		\caption{Behavioral state diagram of the obfuscation tool}
	
	\label{fig:BehavioralStateDiagram}    
\end{figure*}
Now, consider connection $k_{a}'$ which represents intrusive communication $k_a$ executed by employment of non-payload-based obfuscations aimed at modification of its network behavioral properties.
The obfuscations can modify the $P_c$ and $P_s$ packet sets of the original connection $k_a$ by insertion, removal and transformation of the packets.
The modifications of $P_c$ and $P_s$ of the connection $k_a$ can cause alteration of the original features' values $F^a$ to new ones.
Thus, features extracted over $k_{a}'$ are represented by 
\begin{eqnarray}
f(k_{a}') \mapsto F^{a'} = (F_1^{a'}, F^{a'}_2, \ldots, F^{a'}_n)
\end{eqnarray}
and have different values than features $F^{a}$ of the connection $k_{a}$.
Therefore, we conjecture that the likelihood of a correct prediction of $k_a'$-connection's features $F^{a'}$ by the previously assumed classifier $C$ is lower than in the case of connection $k_a$.
Also, we conjecture that classifier $C'$ trained by learning algorithm $A$ on training dataset $D_{tr}'$, containing some obfuscated intrusion instances, will be able to correctly predict higher number of unknown obfuscated intrusions than classifier $C$. 
These assumptions will be evaluated and analyzed later.

\subsection{Obfuscation Tool}
We designed a tool that morphs network characteristics of a TCP connection at network and transport layers of the TCP/IP stack by applying one or a combination of several non-payload-based obfuscation techniques.
Execution of direct communications (non-obfuscated ones) is also supported by the tool as well as capturing network traffic related to a communication. 
The tool is capable of automatic/semi-automatic run and restoring of all modified system settings and consequences of attacks/legitimate communications on a target machine.
After the successful execution of each desired obfuscation on the selected service, the output contains several network packet traces associated with pertaining obfuscations.
The behavioral state diagram of the obfuscation tool is depicted in Figure~\ref{fig:BehavioralStateDiagram}.

\subsection{Description of Data Collection}
We applied the obfuscation tool for a specific set of vulnerable network services and obtained samples of network packet traces related to malicious as well as legitimate communications executed with the employment of particular obfuscations in a virtual network environment. 
Also, we collected network traffic samples of direct attacks for each vulnerable service. 
These network packet traces were passed to a feature extraction process that first identified all TCP connections and then extracted features per each TCP connection.
The collection of these TCP connection-based feature vectors is referred to as dataset, which is analyzed in further machine learning experiments.

\setlength{\tabcolsep}{4pt}    
\begin{table*}[t]
		\footnotesize{
		\begin{center}
						
			\begin{tabular}{p{3.1cm} p{10.0cm} m{0.3cm}}
				\toprule
				\textbf{Technique} & \textbf{Parametrized Instance} & \textbf{ID}\\            
				\toprule
				
				\multirow{3}{4cm}[-0.2cm]{{\textbf{Spread out packets\\in time} }} & {$\bullet$ constant delay: 1s} & {(a)} \\
				
				& {$\bullet$ constant delay: 8s } & {(b)}\\
				
				& {$\bullet$ normal distribution of delay with 5s mean 2.5s standard deviation (25\% correlation)} & {(c)}\\
				\midrule 
				
				\textbf{Packets' loss} & {$\bullet$ 25\% of packets} & {(d)}\\
				\midrule 
				\multirow{3}{4cm}{{\textbf{Unreliable network\\channel simulation}}} & {$\bullet$ 25\% of packets damaged} & {(e)}\\
				
				& {$\bullet$ 35\% of packets damaged} & {(f)}\\
				
				& {$\bullet$ 35\% of packets damaged with~25\% correlation} & {(g)}\\
				\midrule 
				\textbf{Packets' duplication} & {$\bullet$ 5\% of packets} & {(h)}\\
				\midrule 
				\multirow{2}{4cm}[-0.2cm]{\textbf{Packets' order\\modifications}} & {$\bullet$ reordering of 25\% packets;  reordered packets are sent with~10ms delay and 50\%~correlation} & {(i)}\\
				
				& {$\bullet$ reordering of 50\% packets;  reordered packets are sent with~10ms delay and 50\%~correlation} & {(j)}\\
				\midrule 
				\multirow{4}{4cm}{\textbf{Fragmentation}} & {$\bullet$ MTU 1000} & {(k)}\\
				
				& {$\bullet$ MTU 750} & {(l)}\\
				
				& {$\bullet$ MTU 500} & {(m)}\\
				
				& {$\bullet$ MTU 250} & {(n)}\\
				\midrule 
				\multirow{3}{4cm}[-0.5cm]{\textbf{Combinations}} & {$\bullet$ normal distribution delay ($\mu = 10ms$, $\sigma = 20ms$) and 25\% correlation; loss:
					23\% of packets; corrupt: 23\% of packets; reorder:~23\% of packets} & {(o)}\\
				
				& {$\bullet$ normal distribution delay ($\mu = 7750ms$, $\sigma = 150ms$) and 25\% correlation; loss:~0.1\% of packets;
					corrupt:~0.1\% of packets; duplication:~0.1\% of packets; reorder: 0.1\% of packets} & {(p)}\\
				
				& {$\bullet$ normal distribution delay ($\mu = 6800ms$, $\sigma = 150ms$) and 25\% correlation; loss: 1\% of packets;
					corrupt:~1\% of packets; duplication:~1\% of packets; reorder~1\% of packets} & {(q)}\\
				\bottomrule
				
			\end{tabular}		
		\end{center}
	}
	\caption{Experimental obfuscation techniques with parameters and IDs}
	\label{tab:obfuscation-techniques}

\end{table*}
\setlength{\tabcolsep}{1.4pt}     

\subsection{Description of Machine Learning Experiments}        
We performed several classification experiments in order to evaluate the effectiveness of the proposed obfuscation techniques as well as feedback of a classifier having obfuscated data included in its training process.
All of our experiments considered two class prediction, discerning between legitimate and malicious TCP connections. 
Therefore, obfuscated and direct attacks were represented by the same class.
We executed the following experiments:
\begin{compactitem}
	\item For the purpose of finding the best subset of network connection features, we ran the Forward Feature Selection (FFS) method.
	FFS started to run with an empty set of features and in each iteration executing cross validation, it added a new feature contributing by the best improvement of average recall of all classes.
	In order to alleviate the possibility of the selection process becoming stuck in local extremes, we allowed acceptance of one iteration without improvement. 

	\item Considering the selected subset of features, we evaluated evasion resistance of a classifier trained on direct attacks and legitimate traffic only, while testing was performed on the whole dataset including obfuscated attacks.
	
	\item Next, we widened the knowledge of a classifier by adding some obfuscated attacks into the training set and compared its evasion resistance with the previous case.
	
	\item Another experiment tested capability of the classifier to detect unknown obfuscated attacks by customized leave-one-out validation.
	
	\item Finaly, we analyzed the success rate of evasion per particular vulnerable service.	
\end{compactitem}

\section{Implementation}\label{sec:impl}
The proposed obfuscation techniques had been instantiated as part of the obfuscation tool designed and implemented in the Unix environment. 
Parametrized instances of these techniques (introduced in our previous work~\cite{homoliak2016exploitation}) are presented in  Table~\ref{tab:obfuscation-techniques}.
The selection of particular obfuscation techniques was primarily motivated by the need for achieving divergent behavior of obfuscated network attacks as well as by capabilities of Unix OS.
We experimented with various parameters' values with the intention to cover a wide range of divergent behaviors and, moreover, for the case of attacks, preserve the exploitation successful.\footnote{Note that some obfuscations or their combinations may cause an attack to fail, therefore we filtered out such combinations.} 
The methodology presented in this paper allows for a straightforward extension of the proposed obfuscation set.

\subsection{Implementation Notes and Setup}
The obfuscation tool is based on open source tools and is written in the Python and Ruby programming languages. 
For the purpose of an automatic attack execution an utility from \textit{Metasploit} framework was used.
\textit{Tcpdump} tool was chosen to perform network traffic capture between the attacker's machine and vulnerable one. 
Most obfuscations were carried out by Linux \textit{tc} utility and its extension \textit{NetEm}~\cite{hemminger2005netem}, respectively. 
NetEm enabled us to add latency of packets, loss of packets, duplication of packets, reordering of packets, and other outgoing traffic characteristics of the selected network interface. 
The modification of MTU was performed by the Linux utility \textit{ifconfig}. 

\begin{figure*}[t]
	\centering{}    
	\includegraphics[width=10.8cm]{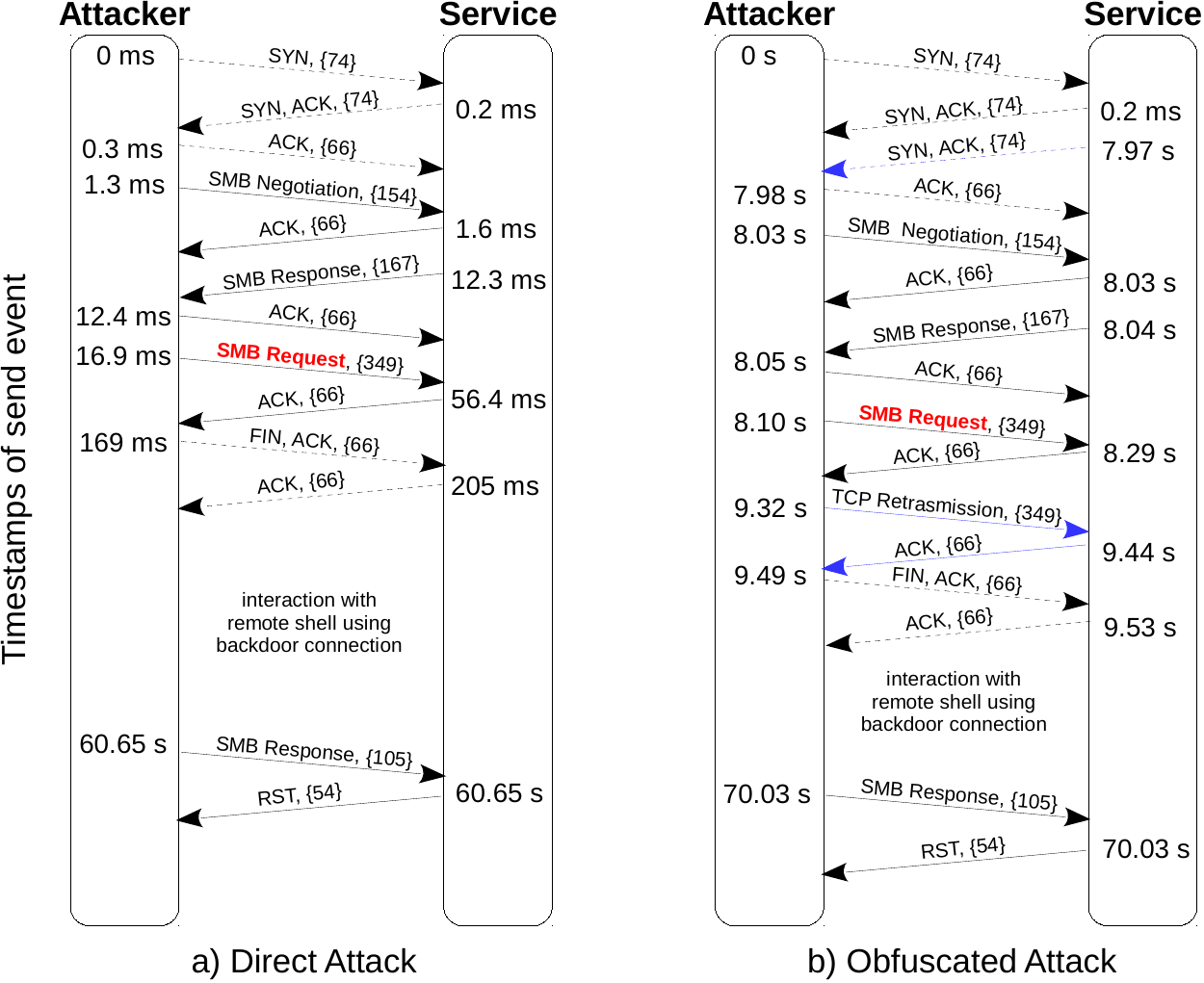}
	\caption[]{Comparison of direct and obfuscated attacks on Samba service}
	\label{fig:TCP-seq-diagram}    
\end{figure*}

We established a virtual network environment for vulnerability exploitation, where all virtual machines (VMs) were configured with private static IP addresses in order to enable easy automation of the whole exploitation process. 
Our testing network infrastructure consisted of the attacker's machine equipped with Kali Linux and vulnerable machines that were running Metasploitable~1,~2,\footnote{\url{https://information.rapid7.com/metasploitable-download.html}} and Windows XP with SP 3.

\subsection{Vulnerable Services}\label{VulnerableServices}
For proof-of-concept purpose, we aimed at selection of vulnerable services with the high severity of their successful exploitation leading to remote shell code execution through an established backdoor communication.
Although there exist plethora of publicly available exploit-codes for contemporary vulnerabilities, the situation with corresponding available vulnerable SW is different due to understandable prevention reasons.
Therefore, we selected older available high-severity vulnerable services that are outdated but may serve as a demonstration of our approach.  
The following listing contains an enumeration of vulnerable services involved in our experiments, complemented by brief description of their exploitation:
%
\begin{compactenum}
	\item \textbf{Apache Tomcat}: 
	-- firstly, a dictionary attack was executed in order to obtain access credentials into the application manager instance.
	Further, the server's application manager was exploited for transmission and execution of malicious code.        
	
	\item \textbf{Microsoft SQL Server}:  
	-- a dictionary attack was employed to obtain access credentials of MSSQL user and then the procedure \texttt{xp\_cmd\-shell} enabling the execution of an arbitrary code was exploited.         
	
	\item \textbf{Samba service}: 
	-- vulnerability in Samba service enabled the attacker of arbitrary command execution, which exploited MS-RPC functionality when configuration \texttt{username map script} was allowed.
	There was no need of authentication in this attack.
	
	\item \textbf{Server service of Windows}: 
	-- the service enabled the attacker of arbitrary code execution through crafted RPC request resulting into stack overflow during path canonicalization.              
	
	\item \textbf{PostgreSQL database}: 
	-- a dictionary attack was executed in order to obtain access credentials  into the PostgreSQL instance.
	Standard PostgreSQL Linux installation had write access to \texttt{/tmp} directory and it could call user defined functions (UDF) that utilized shared libraries located on an arbitrary path (e.g., \texttt{/tmp}). 
	An attacker exploited this fact and copied its own UDF code to \texttt{/tmp} directory and then executed it.
	
	\item \textbf{DistCC service}: 
	-- vulnerability enabled the attacker remote execution of an arbitrary command through compilation jobs that were executed on the server without any permission check. 
	
\end{compactenum}

\setlength{\tabcolsep}{4pt}
\begin{table*}[t]
	\begin{center}		
		\footnotesize{
			\begin{tabular}{>{\raggedleft}p{3.0cm}c>{\centering}p{2.2cm}>{\centering}p{2.0cm}>{\centering}p{2cm}}
				\Xhline{2\arrayrulewidth} \noalign{\smallskip}
				\multirow{2}{3.6cm}{\textbf{~Network Service}} & \multicolumn{4}{c}{\textbf{Count of TCP Connections}} \smallskip \tabularnewline
				\expandafter\cline\expandafter{\expandafter2\string-5\smallskip} & \textbf{Legitimate} & \textbf{Direct Attacks} & \textbf{Obfuscated Attacks} & \textbf{Summary}\tabularnewline             
				\noalign{\smallskip} \Xhline{2\arrayrulewidth} \noalign{\smallskip}
				
				\textbf{Apache Tomcat} & 809 & 61 & 163 & 1033\tabularnewline
				
				\textbf{DistCC} & 100 & 12 & 23 & 135\tabularnewline
				
				\textbf{MSSQL} & 532 & 31 & 103 & 666\tabularnewline
				
				\textbf{PostgreSQL} & 737 & 13 & 45 & 795\tabularnewline
				
				\textbf{Samba} & 4641 & 19 & 44 & 4704\tabularnewline             
				
				\textbf{Server} & 3339 & 26 & 100 & 3465\tabularnewline             
				
				\textbf{Other Legitimate Traffic} & 647 & N/A & N/A & 647\tabularnewline
				
				\midrule
				\textbf{Summary} & 10805 & 162 & 478 & 11445\tabularnewline
				\Xhline{2\arrayrulewidth} 
			\end{tabular}
		}
	\end{center}		
	\caption{Distribution of TCP connection objects in collected dataset}
	\label{tab:Dataset}
\end{table*}
\setlength{\tabcolsep}{1.4pt}

\medskip
An example of a TCP sequence diagram comparing direct and obfuscated attacks on Samba service is depicted in Figure~\ref{fig:TCP-seq-diagram}, where each arrow contains the timestamp of send event and short description with the size of transmitted data.
TCP handshakes and end-shakes are represented by dashed arrows, exploitation of a vulnerability is depicted in red.
The new and different transmissions of obfuscated attack against direct one are depicted in blue, while another difference can be seen in values of transmission time.
These changes were caused by obfuscation $(o)$ that generated a loss of one or more SYN/ACK packets at the 3-way handshake phase, loss of ACK packet after SMB Request and also addition of delay into delivery of all packets.

\subsection{Collected Network Traffic Dataset}\label{NetworkDataset}
We applied our obfuscation tool for automatic exploitation of the enumerated vulnerable services using the proposed obfuscations. 
The captured related malicious network traffic, which we further passed to TCP connection-level feature extractor~\cite{homoliakasnm2013}.
When an exploitation leading to a remote shell was successful, simulated attackers performed simple activities involving various shell commands (such as listing directories, opening and reading files). 
The average number of issued commands was around 10 and text files of up to 50kB were opened/read.
Note that we labeled each TCP connection representing dictionary attacks as legitimate ones due to two reasons: 1.) from the behavioral point of view, they independently appeared just as unsuccessful authentication attempts, which may occur in legitimate traffic as well, 2.) more importantly, we employed ASNM features whose subset involves context of an analyzed TCP connection for their computation -- i.e., ASNM features capture relations to other TCP connections initiated from/to a corresponding service.
On the other hand, legitimate network traffic was collected from two sources:
\begin{compactenum}
	\item Common usage of all previously mentioned services was obtained in an annonymized form, excluding the payload, from a real campus network with accordance to policies in force. 
	Analyzing packet headers, we observed that a lot of expected legitimate traffic contained malicious activity, as many students did not care about up-to-date software.  
	Therefore, we filtered out network connections yielding high and medium severity alerts by signature-based Network Intrusion Detection Systems (NIDS) -- Suricata and Snort -- through Virus Total API~\cite{virusTotalUrl}.
	
	\item The second source represented legitimate traffic simulation in our virtual network architecture and also employed all of our non-payload-based obfuscations for the purpose of partially addressing overstimulation in adversarial attacks against IDS~\cite{corona2013adversarial}, and thus making the classification task more challenging.
	However, only 109 TCP connections were obtained from this stage, which was also caused by the fact that services such as Server and DistCC were hard to emulate.\footnote{Note that additionally to those 109 TCP connections that were explicitly simulated, other 2252 TCP connections from obfuscated dictionary attacks were also considered as legitimate, and thus also helped in achieving a resistance against the overstimulation attacks.}
	Simulation of legitimate traffic was aimed at various \textit{SELECT} and \textit{INSERT} statements when interacting with the database services (i.e., PostgreSQL, MSSQL); several \textit{GET} and \textit{POST} queries to our custom pages as well as downloading of high volume data when interacting with our HTTP server (i.e., Apache Tomcat); and several queries for downloading and uploading small files into Samba share. 
\end{compactenum}
The final dataset is summarized in Table~\ref{tab:Dataset} and is also available from \url{http://www.fit.vutbr.cz/~ihomoliak/asnm/ASNM-NPBO.html}.

\section{Evaluation}\label{DataMiningExperiments}
All machine learning experiments were performed in Rapid Miner Studio~\cite{RapidMiner} using five different classifiers: two with \textit{parametric models} -- Gaussian Na\"ive Bayes and Logistic Regression; and three with \textit{nonparametric models} -- Gaussian Na\"ive Bayes with kernel density estimation, SVM with radial kernel function, and Decision Tree with maximal depth of 10 levels.
Note that parametric models make assumptions about the data, which means that they use a finite set of parameters for modeling the data.
This makes them simple and fast, but on the other hand they are not flexible in modeling of data that do not contain their assumed distribution.
In contrast to them, non-parametric models have no assumptions about the data, and thus they may use unlimited number of parameters. 
The advantage of these models is their flexibility, but on the other hand they may overfit the training data.
Across all of our experiments, network connection features were instantiated by FFS-selected subset of ASNM features (e.g., Table~\ref{FFS-features}), whose full list is available in Appendix~D of~\cite{2016-ihomoliak-thesis}.
Note that some features of ASNM may lead to overfitting of training data due to laboratory conditions of VMs' setup where attacks were executed.
Therefore, such features were removed from the dataset in the preprocessing phase of our experiments and consist of TTL-based features, IP addresses, ports, MAC addresses, occurrence of source/destination host in monitored network.
Considering our current dataset's class distribution, we decided to select 5-fold cross validation, which creates big enough folds for binary classification.
All cross validation experiments have been adjusted to employ stratified sampling during assembling of folds, which ensured equally balanced class distribution of each fold. 

\setlength{\tabcolsep}{4pt}
\begin{table}[t]
		\centering
	\begin{footnotesize}		
		\begin{tabular}{r r r r r}
			\toprule
			\textbf{Classifier} & \textbf{TPR} & \textbf{FPR} & $\mathbf{F_1 ~(\uparrow)}$~~ & \textbf{Avg. Recall}  \\
			\midrule
			
			N. Bayes (kernels)  & 98.15\% & 0.02\% & 98.45\% & 99.07\%  \\
			
			Decision Tree & 95.68\% & 0.09\% & 94.80\% & 97.80\% \\						
			
			SVM  & 82.72\% & 0.01\% & 90.24\% & 91.36\% \\

			Log. Regression & 70.99\% & 0.20\% & 76.92\% & 85.40\% \\

			N. Bayes & 97.53\% & 8.14\% & 26.33\% & 94.70\% \\
			
			\bottomrule
		\end{tabular}
	\end{footnotesize}
	\smallskip
	\caption{Direct attacks and legitimate traffic cross validation}\label{tab:evasion-of-classifiers}
\end{table}	\setlength{\tabcolsep}{1.4pt}	
\setlength{\tabcolsep}{4pt}
\begin{table}[t]
		\centering
	\begin{footnotesize}		
		\subfloat[\label{tab:prediction-obfus}Obfuscated attacks]{
			\begin{tabular}{r r r}
				\toprule
				\textbf{Classifier} & \textbf{TPR $(\uparrow)$} & \textbf{$\mathbf{\Delta}$ TPR} \\
				\midrule
				
				N. Bayes  & 81.80\% & -13.26\% \\				
				
				Log. Regression & 63.18\% & -7.81\% \\

				N. Bayes (kernels)  & 52.30\% & -45.85\% \\
				
				Decision Tree & 36.61\% & -59.07\% \\
				
				SVM  &  15.90\% & -66.82\% \\

				\bottomrule
			\end{tabular}
		}\\
	\subfloat[\label{tab:prediction-all}All attacks]{
			\begin{tabular}{r r r}
				\toprule
				\textbf{Classifier} & \textbf{TPR $(\uparrow)$} & \textbf{$\mathbf{\Delta}$ TPR} \\
				\midrule
				
				N. Bayes  & 86.09\% & -8.97\% \\
				
				Log. Regression & 66.25\% & -4.74\% \\						

				N. Bayes (kernels)  & 64.38\% & -33.77\%\\
				
				Decision Tree &  52.03\% & -43.65\% \\
				
				SVM  & 26.25\% & -56.47\% \\

				\bottomrule
			\end{tabular}
		}
	\end{footnotesize}
		\caption{Prediction of obfuscated/all attacks by classifiers trained without knowledge about obfuscated attacks}
		\label{tab:prediction-obfus-and-all}
\end{table}	\setlength{\tabcolsep}{1.4pt}	
\setlength{\tabcolsep}{4pt}
\begin{table*}
		\centering
	\begin{footnotesize}
		
		\subfloat[\label{tab:prediction-obfus-ffs-dl}FFS DL features]{
			\begin{centering}
				\protect\centering{}
				\begin{tabular}{r r r r r r r}
					\toprule
					\textbf{Classifier} & \textbf{TPR}~ & \textbf{FPR}~ & \specialcell{~~$\Delta$\\\textbf{TPR}} & \specialcell{~~$\Delta$\\ \textbf{FPR}} & $\mathbf{F_1} (\uparrow)$~~~ & \textbf{\specialcell{~Avg.\\Recall}} \\
					\midrule
					
					N. Bayes (kernels)  & 93.28\% & 0.73\% & +28.90\% &	+0.71\%  & 90.73\% & 96.28\%  \\
					
					SVM  &  80.31\% & 0.05\% & +54.06\% & +0.04\% & 88.70\% & 90.13\% \\
					
					Log. Regression & 74.69\% & 0.33\% & +20.00\% & +0.13\% & 82.85\% & 87.18\% \\

					Decision Tree & 67.34\% & 0.36\% &  +15.31\% &	+0.27\% & 77.65\% & 83.49\% \\

					N. Bayes & 60.31\% & 1.87\% & -34.07\% & -6.27\% &	62.87\% & 79.22\% \\

					\bottomrule
				\end{tabular}\protect
			\end{centering}
		}
		
		\subfloat[\label{tab:prediction-obfus-ffs-dol}FFS DOL features]{
			\begin{centering}
				
				\protect\centering{}
				\begin{tabular}{r r r r r r r}
					\toprule
					\textbf{Classifier} & \textbf{TPR}~ & \textbf{FPR}~ & \specialcell{~~$\Delta$\\\textbf{TPR}} & \specialcell{~~$\Delta$\\ \textbf{FPR}} & $\mathbf{F_1} (\uparrow)$~~~ & \textbf{\specialcell{~Avg.\\Recall}} \\
					\midrule
					
					SVM & 99.53\% & 0.13\% & +73.28\% & +0.12\% & 98.68\% & 99.70\% \\
					
					Decision Tree & 98.44\%	& 0.19\% & +46.41\% & +0.10\% & 97.60\% & 99.13\% \\
					
					N. Bayes (kernels)  & 98.75\% & 0.99\% & +34.37\% & +0.97\% & 91.66\%	& 98.88\% \\																
					
					Log. Regression  & 97.50\% & 1.68\% & +31.25\% & +1.48\% & 86.37\% & 97.91\%  \\

					N. Bayes  & 98.59\% & 3.75\% & +4.21\% & -4.39\% & 75.30\% & 97.42\%\\								
					
					\bottomrule
				\end{tabular}\protect
			\end{centering}
		}
			\caption{Whole dataset cross validation}
			\label{tab:bi-all-cross}
			\vspace{-0.5cm}
	\end{footnotesize}
	\end{table*}	
\setlength{\tabcolsep}{1.4pt}

\subsection{Forward Feature Selection}	
The experiment consisted of two executions of the FFS per each classifier. 
In each of executions, we optimized a few important parameters of the classifiers using grid approach.
In the cases of both Na\"ive Bayes classifiers, we enabled Laplace correction in order to prevent models from high influence by zero probabilities of some values, and moreover in the kernel-based version we optimized the bandwidth of kernels.
In SVM, we optimized: 1) parameter C that represents trade-off between a soft and hard boundary of the hyperplane and 2) parameter gamma of the Gaussian radial kernel that influences the variance of the Gaussian kernel.
The regularization parameter lambda was optimized in the case of Logistic Regression -- the parameter controls overfitting of the model at the expense of incorporating the bias.
And finally, in the case of the decision tree, we used gain ratio as a criterion for selection of attributes for splitting, while we optimized minimal gain required for splitting, which controls the number of splits.

The first execution set of FFS took as input just legitimate traffic and direct attack entries, and represented the case where intrusion detection classifiers were trained without knowledge about obfuscated attacks.
We denote the selected features as \textbf{FFS DL} (\textbf{D}irect + \textbf{L}egitimate).
The second execution set took as input the whole dataset of network traffic -- consisting of legitimate traffic, direct attacks as well as obfuscated ones, and thus represented the case where classifiers were aware of obfuscated attacks.
Here, we denote the selected features as \textbf{FFS DOL} (\textbf{D}irect + \textbf{O}bfuscated + \textbf{L}egitimate).
We assume FFS DL features set as less informed (and thus less tuned) than FFS DOL features, therefore when FFS DL features are used, we assume that classifiers do not have knowledge about obfuscated attacks, while FFS DOL features are used when we assume the opposite case.
Both FFS-selected feature sets are presented in Table~\ref{FFS-features} of Appendix (columns FFS DOL and FFS DL).

\subsection{Evasion of Intrusion Detection Classifiers}\label{subsection:binominalExI} 
A 5-fold cross validation was performed using direct attacks with legitimate traffic using FFS DL features.
The performance measures of the classifiers validated by cross validation are shown in Table~\ref{tab:evasion-of-classifiers}.
Then the classifiers trained on all direct attacks and legitimate traffic instances were applied for the prediction of the obfuscated attacks and all attacks, respectively (see Table~\ref{tab:prediction-obfus-and-all}). 
Here TPRs were deteriorated for all classifiers, which means that some obfuscated attacks were successful -- they were predicted as legitimate traffic, and thus caused evasion of the classifiers. 

Note that in the case of direct attacks and legitimate traffic cross validation, non-parametric classifiers achieved better performance than parametric classifiers, while in the case of obfuscated attacks non-parametric classifiers were more significantly deteriorated in TPR, which was caused by their property of overfitting known data.

\subsection{Widening the Knowledge of the Classifiers}
In order to improve the resistance of the classifiers against evasions, we widened their knowledge about different mixtures of obfuscated attack instances,
which was accomplished by random 5-fold cross validation of the whole dataset. 
In this experiment, it is justified to use FFS DOL features that consider knowledge about obfuscated attacks for updating not only the model of a classifier but also underlying feature set. 
Additionally, we show the results with FFS DL features, which consider updating model only.
The results of this experiment are shown in Table~\ref{tab:bi-all-cross}.
Comparing against the results from the previous experiment (see FPRs from Table~\ref{tab:evasion-of-classifiers} and TPRs from Table~\ref{tab:prediction-obfus-and-all}b), 
most of the classifiers were significantly improved in TPR, while FPR was deteriorated only slightly.
This confirms the fulfilled assumption that the classifiers trained with knowledge about some obfuscated attacks are able to detect the same and similar obfuscated attacks.
The only exception is the Gaussian Na\"{i}ve Bayes classifier that updates model only, not the underlying feature set (Table~\ref{tab:bi-all-cross}a).
Here is important to note that this classifier makes strong assumptions about the modeled data and when we searched for the optimal feature set with direct and legitimate traffic (FFS DL), it was unable to further optimize FPR, which remained high in contrast to other classifiers.
Therefore, when obfuscated attacks were added into the cross validation, the classifier was unable to use the same features and the same strong assumptions about the original data for fitting the different data.
However, in the case of updating the feature set (Table~\ref{tab:bi-all-cross}b), both TPR and FPR of the classifier were improved. 

\begin{table*}
		\centering{}
	\footnotesize{		
		\subfloat[\label{tab:unknown-obfuscations-single}Per instance]{\protect\begin{centering}
				\footnotesize{
					\begin{tabular}{>{\centering}p{2.3cm}>{\centering}m{1.8cm}>{\centering}m{1.3cm}>{\centering}m{1.3cm}>{\centering}m{1.3cm}>{\centering}m{1.3cm}>{\centering}m{1.3cm}>{\centering}m{1.3cm}}										
						\Xhline{3\arrayrulewidth} \noalign{\smallskip}

						\multirow{2}{2.0cm}[-0.4cm]{\textbf{~Unknown\\\hspace{0.1cm}Instance of\\Obfuscation\\~Technique}}  & 
						\multirow{2}{1.6cm}[-0.4cm]{\textbf{~~The\\Number\\~~~~of\\Samples}} &
						\multicolumn{6}{c}{\textbf{Ratio of Correctly Detected Samples}} \tabularnewline
						\expandafter\cline\expandafter{\expandafter3\string-8\smallskip}
						
						& & 
						\rotatebox[origin=c]{90}{\textbf{\specialcell{N. Bayes\\(Kernels)}}} &
						\rotatebox[origin=c]{90}{\textbf{N. Bayes}} &
						\rotatebox[origin=c]{90}{\textbf{\specialcell{Decision\\~~Tree}}} &
						\rotatebox[origin=c]{90}{\textbf{SVM}} &
						\rotatebox[origin=c]{90}{\textbf{\specialcell{~~Logistic\\Regression}}} &
						\rotatebox[origin=c]{90}{\textbf{\specialcell{~Average\\~~~of All\\Classifiers}}}
						\tabularnewline
						\noalign{\smallskip} \Xhline{1\arrayrulewidth} \noalign{\smallskip}
						
						(a) & 28 & 100.00\% & 100.00\% & 100.00\% & 100.00\% & 100.00\% & 100.00\% \tabularnewline 
						
						(b) & 22 & 100.00\% & 100.00\% & 100.00\% & 100.00\% & 100.00\% & 100.00\% \tabularnewline 
						
						(j) & 27 & 100.00\% & 100.00\% & 100.00\% & 100.00\% & 100.00\% & 100.00\% \tabularnewline 
						
						(p) & 33 & 100.00\% & 100.00\% & 100.00\% & 100.00\% & 100.00\% & 100.00\% \tabularnewline 
						
						(c) & 30 & 96.67\% & 100.00\% & 100.00\% & 100.00\% & 100.00\% & 99.33\% \tabularnewline 
						
						(i) & 27 & 100.00\% & 100.00\% & 100.00\% & 96.30\% & 100.00\% & 99.26\% \tabularnewline 
						
						(e) & 26 & 100.00\% & 100.00\% & 96.15\% & 100.00\% & 100.00\% & 99.23\% \tabularnewline 
						
						(g) & 26 & 100.00\% & 100.00\% & 100.00\% & 100.00\% & 96.15\% & 99.23\% \tabularnewline 
						
						(d) & 30 &  100.00\% & 100.00\% & 96.67\% & 96.67\% & 100.00\% & 98.67\% \tabularnewline 
						
						(h) & 30 & 100.00\% & 100.00\% & 96.67\% & 96.67\% & 100.00\% & 98.67\% \tabularnewline 
						
						(q) & 35 & 100.00\% & 97.14\% & 97.14\% & 97.14\% & 100.00\% & 98.29\% \tabularnewline 
						
						(m) & 27 & 92.59\% & 100.00\% & 92.59\% & 100.00\% & 100.00\% & 97.04\% \tabularnewline 
						
						(o) & 28 & 100.00\% & 92.86\% & 92.86\% & 100.00\% & 96.43\% & 96.43\% \tabularnewline 
						
						(f) & 28 & 92.86\% & 96.43\% & 92.86\% & 100.00\% & 96.43\% & 95.71\% \tabularnewline 
						
						(l) & 27 & 81.48\% & 100.00\% & 92.59\% & 100.00\% & 100.00\% & 94.81\% \tabularnewline 
						
						(k) & 27 & 66.67\% & 100.00\% & 100.00\% & 100.00\% & 100.00\% & 93.33\% \tabularnewline 
						
						(n) & 27 & 70.37\% & 77.78\% & 48.15\% & 55.56\% & 74.07\% & 65.19\% \tabularnewline 
						
						\noalign{\smallskip} \Xhline{1\arrayrulewidth} \noalign{\smallskip}
						\textbf{Average} & & 94.15\% & \textbf{97.89\%} & 94.45\% & 96.61\% & \textbf{97.83\%} & \tabularnewline 
						\textbf{Std. Dev.} & & $\pm$10.81\% & \textbf{$\pm$5.54\%} & $\pm$12.30\% & $\pm$10.68\% & \textbf{$\pm$6.29\%} & \tabularnewline 					
						
						\Xhline{2\arrayrulewidth}
					\end{tabular}
				}
				
				\protect
				\par\end{centering}
		}
		
		\subfloat[\label{tab:unknown-obfuscations-group}Per technique]{\protect\begin{centering}
				\footnotesize{
					\begin{tabular}{>{\raggedleft}p{2.3cm}>{\centering}m{1.8cm}>{\centering}m{1.3cm}>{\centering}m{1.3cm}>{\centering}m{1.3cm}>{\centering}m{1.3cm}>{\centering}m{1.3cm}>{\centering}m{1.3cm}}										
						\Xhline{3\arrayrulewidth} \noalign{\smallskip}

						\multirow{2}{2.0cm}[-0.4cm]{\textbf{~Unknown\\Obfuscation\\~Technique}}  & 
						\multirow{2}{1.6cm}[-0.4cm]{\textbf{~~The\\Number\\~~~~of\\Samples}} &
						\multicolumn{6}{c}{\textbf{Ratio of Correctly Detected Samples}} \tabularnewline
						\expandafter\cline\expandafter{\expandafter3\string-8\smallskip}
						
						& & \rotatebox[origin=c]{90}{\textbf{\specialcell{N. Bayes\\(Kernels)}}} &
						\rotatebox[origin=c]{90}{\textbf{N. Bayes}} &
						\rotatebox[origin=c]{90}{\textbf{\specialcell{Decision\\~~Tree}}} &
						\rotatebox[origin=c]{90}{\textbf{SVM}} &
						\rotatebox[origin=c]{90}{\textbf{\specialcell{~~Logistic\\Regression}}} &
						\rotatebox[origin=c]{90}{\textbf{\specialcell{~Average\\~~~of All\\Classifiers}}}
						\tabularnewline
						\noalign{\smallskip} \Xhline{1\arrayrulewidth} \noalign{\smallskip}
						
						(a, b, c) & 80 & 98.75\% & 100.00\% & 100.00\% & 100.00\% & 100.00\% & 99.75\% \tabularnewline
						
						(i, j) & 54 & 100.00\% & 100.00\% & 100.00\% & 94.44\% & 100.00\% & 98.89\% \tabularnewline
						
						(d) & 30 &  100.00\% & 100.00\% & 96.67\% & 96.67\% & 100.00\% & 98.67\% \tabularnewline
						
						(h) & 30 & 100.00\% & 100.00\% & 96.67\% & 96.67\% & 100.00\% & 98.67\% \tabularnewline
						
						(o, p, q) & 96 & 100.00\% & 96.88\% & 96.88\% & 96.88\% & 98.96\% & 97.92\% \tabularnewline
						
						(e, f, g) & 80 & 97.50\% & 98.75\% & 95.00\% & 100.00\% & 97.50\% & 97.75\% \tabularnewline
						
						(k, l, m, n) & 108 & 75.93\% & 92.59\% & 83.33\% & 61.11\% & 93.52\% & 81.30\% \tabularnewline

						\noalign{\smallskip} \Xhline{1\arrayrulewidth} \noalign{\smallskip}
						\textbf{Average} & & 96.03\% & \textbf{98.32\%} & 95.51\% & 92.25\% & \textbf{98.57\%}	&							
						\tabularnewline 
						\textbf{Std. Dev.} & & $\pm$8.91\% & \textbf{$\pm$2.78\%} & $\pm$5.68\% & $\pm$13.87\% & \textbf{$\pm$2.41\%} & 
						\tabularnewline
						
						\Xhline{2\arrayrulewidth}
					\end{tabular}
				}
				
				\protect
				\par\end{centering}
		}
	}
	\caption{Ratios of correctly detected unknown obfuscated attacks}
	\label{tab:unknown-obfuscations}
\end{table*}    
\setlength{\tabcolsep}{1.4pt}    
\begin{table*}
	\centering{}
	\footnotesize{
		
		\begin{tabular}{>{\raggedleft}p{2.3cm}>{\centering}m{1.8cm}>{\centering}m{1.3cm}>{\centering}m{1.3cm}>{\centering}m{1.3cm}>{\centering}m{1.3cm}>{\centering}m{1.3cm}>{\centering}m{1.3cm}}										
			\Xhline{3\arrayrulewidth} \noalign{\smallskip}

			\multirow{2}{2.0cm}[-0.8cm]{\textbf{Vulnerable\\~~Service}}  & 
			\multirow{2}{1.6cm}[-0.4cm]{\textbf{~~The\\Number\\~~~~of\\Samples}} &
			\multicolumn{6}{c}{\textbf{Ratio of Successfully Evaded Samples}} \tabularnewline
			\expandafter\cline\expandafter{\expandafter3\string-8\smallskip}
			
			& & \rotatebox[origin=c]{90}{\textbf{\specialcell{N. Bayes\\(Kernels)}}} &
			\rotatebox[origin=c]{90}{\textbf{N. Bayes}} &
			\rotatebox[origin=c]{90}{\textbf{\specialcell{Decision\\~~Tree}}} &
			\rotatebox[origin=c]{90}{\textbf{SVM}} &
			\rotatebox[origin=c]{90}{\textbf{\specialcell{~~Logistic\\Regression}}} &
			\rotatebox[origin=c]{90}{\textbf{\specialcell{~Average\\~~~of All\\Classifiers}}}
			\tabularnewline
			\noalign{\smallskip} \Xhline{1\arrayrulewidth} \noalign{\smallskip}
			
			Apache & 163 & 55.21\% & 4.91\% & 55.83\% & 88.34\% & 93.87\% & 59.63\% \tabularnewline
			
			PostgreSQL & 45 & 66.67\% & 0.00\% & 88.89\% & 100.00\% & 0.00\% & 51.11\% \tabularnewline 
			
			MSSQL & 103 & 18.45\% & 23.30\% & 71.84\% & 100.00\% & 12.62\% & 45.24\% \tabularnewline 
			
			Samba & 44 & 70.45\% & 0.00\% & 50.00\% & 100.00\% & 2.27\% & 44.55\% \tabularnewline 
			
			DistCC & 23 & 39.13\% & 17.39\% & 95.65\% & 47.83\% & 17.39\% & 43.48\% \tabularnewline 
			
			Server & 100 & 49.00\% & 0.00\% & 54.00\% & 44.00\% & 5.00\% & 30.40\% \tabularnewline

			\noalign{\smallskip} \Xhline{1\arrayrulewidth} \noalign{\smallskip}
			\textbf{Average} & & 49.82\% & \textbf{7.60}\% & 69.37\% & 80.03\% & \textbf{21.86\%} &							
			\tabularnewline 
			\textbf{Std. Dev.} & & 19.17\% & 10.23\% & 19.35\% & 26.84\% & 35.88\% & 
			\tabularnewline
			
			\Xhline{2\arrayrulewidth}
		\end{tabular}
	}
	\smallskip
	\caption{Successfully obfuscated attacks (evasions) per service}
	\label{tab:evasions-per-service}
\end{table*}    
\setlength{\tabcolsep}{1.4pt}

\subsection{Detection of Unknown Obfuscated Attacks}
For the purpose of explicitly testing the classifiers' capability to detect new kinds of obfuscated attacks, we performed customized leave-one-out validation using FFS DOL features, where the classifier was step-by-step trained on all permutations of the whole dataset excluding only obfuscated attack samples created by a single obfuscation technique, or its instance, respectively; while it was validated on the excluded part of the dataset. 
Table~\ref{tab:unknown-obfuscations} presents ordered ratios of correctly detected unknown obfuscated attacks per obfuscation technique as well as per its instance.
Comparing detection performance of unknown obfuscated attacks, either per instance or per obfuscation technique, we concluded that in most of the obfuscations, there were achieved high detection rates that indicate the acceptable resistance of the obfuscations-aware classifiers against unknown obfuscated attacks.
The only exceptions are obfuscation techniques that modify MTU. 
This can be explained by the fact that the majority of the features employed in our experiments is mostly sensitive to packet lengths, which are influenced by fragmentation-based obfuscations.	
This phenomenon is more significant in the cases of non-parametric classifiers due to their property of overfitting the training data. 
In general, parametric classifiers were more successful in detecting unknown obfuscated attacks and their correct predictions were more stable than in non-parametric classifiers. 
However, in the case of one parametric classifier -- Gaussian Na\"{i}ve Bayes -- we also have to take into account its worse FPR in comparison to Logistic Regression (shown in Table~\ref{tab:evasion-of-classifiers}).

\subsection{Successful Evasions per Service}\label{SummaryObfuscationTechniques}
This experiment compares and analyzes success rate of evasion per vulnerable service. 
The results presented here originate from a binary classification experiment in which the classifier was trained without obfuscated attacks and validated on the whole dataset (Table~\ref{tab:prediction-obfus-and-all}) using FFS DL features. 
The obfuscations are considered successful if they are predicted as legitimate traffic; the situation represents evasion case.
Ordered ratios of successfully obfuscated attacks per service are present in Table~\ref{tab:evasions-per-service}.
The minimum achieved ratios of attacks that evaded detection are shown in bold and belong to parametric classifiers, which was already mentioned above.
Most successful obfuscated attacks are those exploiting Apache service.
From a detailed analysis of this service, we found out that instances of direct attacks had very flat value distribution of many features in comparison to other direct attacks.
Examples of such features are the standard deviation of inbound and outbound packet sizes of the connection, and other features dependent on the packets' length variability.
Therefore, obfuscated attacks caused more variability of the features that were in many cases similar to legitimate traffic. 
On the other hand, in the cases of Server, many features of the direct attacks were more divergent across their instances, and thus obfuscations contributed to the divergence only in a low scale.
Therefore, most of the obfuscated attacks had similar characteristics like direct ones, which enabled their detection.

\section{Discussion}

\paragraph{\textbf{Impact of Obfuscations on Feature Divergence.}}
To assess the impact of proposed obfuscations on the divergence of ASNM features, we compared values of each FFS-selected feature of obfuscated attacks with feature values obtained from direct attacks executed with the same exploit.	
In other words, we quantified the change that obfuscations bring by computing the ratio of divergent single-feature obfuscated attacks against the closest single-feature direct attack using the same exploit on the same service.	
We found out that this ratio (averaged per all FFS DL features) is higher than $55\%$ in the case of all classifiers, which can be viewed as the proposed obfuscations were able to influence the majority of features in obfuscated attacks.
The example of this ratio computed per each input feature of Gaussian Na\"ive Bayes classifier is presented in Table~\ref{FFS-features} (the last column).
Note that in the case of FFS DOL features, the average of this ratio per-feature is lower ($66.33\%$) than in the case of FFS DL features ($72.80\%$), as FFS DOL was aware of obfuscations and thus selected more obfuscation-resistant features.

\vspace{-0.3cm}
\paragraph{\textbf{Retraining.}}
Although we had demonstrated that our adversarial classification approach to network intrusion detection can detect unknown obfuscated attacks with high performance, it is still possible to design and apply unknown network connection morphing techniques to bypass the detection.
In order to keep this performance as high as possible, retraining of the classifier should be performed each time a new form of obfuscation is known to occur. 
However, such retraining of \textit{generative classifiers} (both Na{\"i}ve Bayes classifiers) relates to sub-model of the malicious class only, 
and therefore is faster than retraining monolithic models of \textit{discriminative classifiers} such as SVM, decision tree, and logistic regression, where the whole model incorporating both classes has to be retrained.
This favors generative classifiers over discriminative when a frequent update of a model is required.
On the other hand, retraining of the legitimate sub-model of generative classifiers should be also performed once in a while in order to ensure that all new manners of using particular services are captured.
Next aspect related to fast retraining of classifiers is whether they can be retrained with preserving the feature set and still provide high performance. 
From this point of view, we have found Na\"{i}ve Bayes with Gaussian kernels as the most convenient classifier (see results in Table~\ref{tab:bi-all-cross}a).

\vspace{-0.3cm}
\paragraph{\textbf{Extension of Obfuscations.}}
Our obfuscations are not exhaustive but cover a wide range of network connection morphing possibilities that can influence the detection performance of a non-payload-based intrusion detection classifier. 
On the other hand, the methodology presented in this paper allows for a straightforward extension of the obfuscations.

\vspace{-0.3cm}
\paragraph{\textbf{High Rate of Attacks.}}
Our dataset has the ratio of malicious to legitimate connections equal to $5.9\%$, while in practice this ratio is usually several orders of magnitude lesser.
Although an arbitrary value of this ratio does not distort the performance of the classifier when correct performance measure is chosen (e.g., $\varDash{F_1-measure}$, average recall of classes), it might impact the accuracy of modeling the legitimate class whose high volume occurred in practice can result in high divergence of data, which might not be captured by models built from our dataset in sufficient manner. 
Therefore in practice, classifiers would require much more legitimate data than in our dataset.

\paragraph{\textbf{Normalizers.}}
If we would assume the existence of optimal network normalizer that would be able to completely eliminate the impact of proposed non-payload-based obfuscation techniques, then
these obfuscation techniques would be useless. 
Nevertheless, if such optimal network normalizer
would exist, then it would be still prone to state holding and CPU overload attacks~\cite{dharmapurikar2005robust},~\cite{papadogiannakis2012tolerating},~\cite{singh2006stateless}.
Contrary, if we would not assume network normalizer as part of our system, then non-payload-based obfuscation techniques might be employed as \textit{training data driven approximation of network normalizer} that would not be prone to previously mentioned attacks.
The situation can be demonstrated by our binary classification experiments (Section~\ref{subsection:binominalExI}).
Consider intrusion detection classifier validated on direct attacks and legitimate traffic whose average recall is higher than $90\%$ for each classifier (Table~\ref{tab:evasion-of-classifiers}). 
Here training and testing data of the classifier were built upon normalized malicious network traffic represented by direct attacks.
Then, the model trained on the direct attacks and legitimate traffic was applied to prediction of the obfuscated attacks.
In this case, obfuscated attacks may represent un-normalized malicious network traffic, and thus the classifier achieved worse performance than in the previous case: TPR  was significantly decreased while FPR was preserved from the previous step (Table~\ref{tab:prediction-obfus-and-all}a).
In order to alleviate negative performance impact of un-normalized malicious network traffic (represented by obfuscated attacks) on our system, we can include obfuscated attacks in the training process of the classifier. 
This case is interpreted by performance measured contained in Table~\ref{tab:bi-all-cross}b. 
There was achieved average recall over $97\%$ for each classifier, primarily thanks to significant improvement of TPR in most of the classifiers. 
Thus, as an alternative outcome of our work, a network normalizer element may be omitted from classification-based intrusion detection infrastructure and can be approximated by appropriate training data.

\section{Related Work \label{SOA}} 		
Using taxonomy of attacks against learning systems~\cite{barreno2010security}, we categorize our approach to the class of \textit{exploratory indiscriminate} attacks \textit{violating integrity via false negatives}.
The same class of attacks was addressed for example in the field of \textit{spam filtering}~\cite{Biggio-One-and-a-Half-Class}, \cite{lowd2005adversarial}, \textit{malware detection}~\cite{Biggio-One-and-a-Half-Class}, \textit{payload-based anomaly intrusion detection}~\cite{fogla2006polymorphic}, \cite{tan2002undermining}, \textit{automatic speech recognition}~\cite{carlini2018audio}, etc. 
However, to the best of our knowledge, this class of attacks had not been studied in the \textit{non-payload-based intrusion detection} yet, including anomaly-based and classification-based approaches.
Further, we aim at related work of evasive adversarial attacks against IDS, and we divide it into payload-based and non-payload-based approaches, plus their combination.
Additionally, papers dealing with network traffic normalization are also described.

\vspace{-0.4cm}
\paragraph{\textbf{Payload-based Evasions.}}
The first work dealing with payload-based evasions is described in~\cite{puppy1999whiskers} and presents a tool called Whisker.  
The author aims at anti-intrusion detection tactics by performing mutations of the HTTP request in a way that a web server is able to understand the request, but intrusion detection systems can be confused.
Vigna et al.~\cite{vigna2004testing} proposed a framework generating exploit mutations to change the appearance of a~malicious payload bypassing detection of NIDS.
The proposed framework was evaluated on two well-known signature-based NIDSs -- Snort and RealSecure. 
A similar approach was proposed by Fogla et al.~\cite{fogla2006polymorphic} in their polymorphic blending attacks that change the payload of a network worm in order to look normal, and thus effectively evade a byte frequency-based anomaly NIDS. 
Other approaches use many different techniques for evading detection by changing the payload, e.g., obfuscation techniques such as malware morphism~\cite{you2010MalObfTecBriSur_3} and other attack tactics against IDSs~\cite{corona2013adversarial}. 
All of these adversarial approaches are similar to our approach, but in contrast they deal only with evasions of payload-based NIDSs.

\vspace{-0.4cm}
\paragraph{\textbf{Non-payload-based Evasions.}}
Previous methods can evade payload-based NIDS systems primarily by morphing the payload, but do not need to be efficient against non-payload-based network intrusion detectors, which are most sensitive on the attack morphing at the network and transport layers of the TCP/IP stack.
Fragroute~\cite{ptacek1998insertion} is a tool that was written to test intrusion detection systems by using simple ruleset language enabling interception and modification of egress traffic with minimal support for randomized or probabilistic behavior.
Fragroute implements three classes of attacks -- insertion, evasion, and denial of service. 
AGENT~\cite{rubin2004automatic} uses several methods of altering network traffic by packet splitting, duplicate insertions, etc.  
Watson et al.~\cite{watson2004protocol} proposed a method called \textit{Protocol Scrubbing} that represents active mechanisms for transparent removing of network attacks from protocol layers in order to allow passive IDS systems to operate correctly against evasion techniques. 
Wright et al.~\cite{wright2009traffic} proposed thwarting of network traffic classifiers by optimally morphing one class of traffic to look like another class with respect to a given set of features, while they employ padding or splitting the data into smaller parts.
This is similar to our approach, but in contrast the authors aim at network traffic classification in general, rather than intrusion detection. 
An example of evasion that deals with tunneling of malicious network traffic in payload of HTTP(S) protocol is presented in~\cite{homoliak:NBAofObfNetVul} and~\cite{homoliak:ChofBOAinHTTP}.
Although the authors do evasion of payload-based NIDS as well, the main focus of the work is the evasion of non-payload-based NIDS.

\paragraph{\textbf{Combinations.}}
Evasions based on modifications at each of the application, transport and network layers of the TCP/IP stack are described in~\cite{cheng2012evasion} and~\cite{juan2008tool}.
Cheng et al.~\cite{cheng2012evasion} described general evasion techniques and
examined the detection performance of signature-based NIDS when performing mutation of known attacks.
Juan et al. described framework called \textit{idsprobe}~\cite{juan2008tool} that is intended for evasion-resilience testing of NIDSs. 
Idsprobe can perform offline as well as live evasion test cases, while it supports payload-based and non-payload-based modifications of network attacks according to predefined transformation profiles.  
The authors of idsprobe differ from our approach in two points: 1.) they aim at payload-based intrusion detection, 2.) their test cases involve evasions at application layer of ISO/OSI.

\vspace{-0.3cm}
\paragraph{\textbf{Normalizers.}}
In order to answer non-payload-based evasions of NIDS, the concept of network traffic normalizer was introduced by Handley et al.~\cite{handley2001network}.
The authors proposed the implementation of normalizer called \textit{norm}. 
Norm performs normalizations of ambiguities in the TCP traffic stream that can be seen by NIDS. 
How\-ever, introducing a network normalization brought problems related to platform dependent semantic of network ambiguities interpretation as well as throughput reduction. 
These problems lead Shankar et al.~\cite{shankar2003active} to introduce the concept and implementation of \textit{Active Mapping}, which eliminates them with minimal runtime cost by building profiles of the network topology including the TCP/IP policies of hosts on the network.
A NIDS may then use the host profiles to disambiguate the interpretation of the network traffic on a per-host basis. 
However, because of the shortcomings of network normalizers, their usage in a network can result in side-effects and can even be prone to various attacks, e.g., state holding, and CPU overload~\cite{dharmapurikar2005robust},~\cite{papadogiannakis2012tolerating},~\cite{singh2006stateless}.

\section{Conclusion}\label{Conclusion}
The motivation behind our work is to strengthen non-payload-based intrusion detection classifiers in an attack-independent way, assuming an adversary who can massively mutate known exploits to attack huge amount of targets.
With this in mind, we executed remote attacks and legitimate communications on selected vulnerable network services while utilizing various non-payload-based obfuscation techniques based on NetEm and MTU modifications with the intention to make behavioral characteristics of the attacks being similar to those of legitimate traffic, and thus cause evasion of our experimental non-payload-based intrusion detection classifiers. 
The summary of the presented results revealed non-payload-based obfuscation techniques as partially successful in evading detection by five classifiers (two parametric and three non-parametric), which were  trained without prior knowledge about them. 	
On the other hand, if some of the obfuscated attacks were included in the training process of the classifiers, then they were able to detect other unknown obfuscated attacks with high performance. 
From the practical point of view, we discussed requirements on fast retraining of classifiers, where we identified Na\"{i}ve Bayes classifier with Gaussian kernels as the most convenient one due to its capability to update the model of a single class independently of another class, and also it does not need to replace the feature set and still can provide a high performance.
Note that we do not envision to use our obfuscation-aware non-payload-based classifiers as an independent security solution but as a complementary part of existing solutions, such as to misuse-based and anomaly-based intrusion detection systems that perform deep packet inspection.

In our future work, we plan to perform experiments with existing implementations of network normalizers as well as verify the effect of non-payload-based obfuscation techniques on wider spectrum of vulnerabilities.
Another option is to explore impact of proposed obfuscations on communication between bots and C\&C server.

\section*{Acknowledgment} 
The work was supported by the IT4Innovations Excellence in Science project LQ1602, the project Reliability and Security in IT (FITS-14-2486), and The Ministry of Education, Youth and Sports from the National Program of Sustainability (NPU II); project IT4Innovations excellence in science - LQ1602; and Aggregated Quality Assurance for Systems (AQUAS) -- 8A17001.
This research was also supported by ST Electronics and National Research Foundation (NRF), Prime Minister's Office Singapore, under Corporate Laboratory @ University Scheme (Programme Title:  STEEInfosec -- SUTD Corporate Laboratory). 
We would like to thank the Virus Total team for providing access to premium API as well as our colleague Matej Gregr for help in provisioning of anonymized network packet traces.

\bibliographystyle{EAI-num}
\bibliography{article}

\begin{appendices}

\section{Utilized ASNM Features}		

\setlength{\tabcolsep}{2.0pt}
\begin{table*}
	\centering
	\footnotesize{			
		\hspace{-1.0cm}\begin{tabular}{@{}l p{0.05cm} p{9.1cm} c c c c c c@{}}
			\toprule
			
			\textbf{Feature ID} & & \multicolumn{1}{c}{\textbf{Description}} & &  \textbf{\specialcell{FFS\\DOL}}  & 
			\textbf{\specialcell{FFS\\~DL}}  & &
			\textbf{\specialcell{Ratio of Obfuscated\\~~~Attacks Having\\~~Divergent Values\\~~~~of a Feature\\~~in Comparison to\\~~~~Direct Attacks\\~~~~Executed with\\~~the Same Exploit}}
			
			\\                   
			\midrule

			SigPktLenOut   &    & $\bullet$ Std. deviation of outbound (client to server) packet sizes. & & X & X & & 99.44 \% \\    
			
			MeanPktLenIn   &    & $\bullet$ Mean of packet sizes in inbound traffic of a connection. & & X & X & & 95.01 \% \\
			
			CntOfOldFlows &    & $\bullet$ The number of mutual connections between client and server, which started up to 5 minutes before start of an analyzed connection. & & X & X & & 35.37 \% \\
			
			CntOfNewFlows &    & $\bullet$ The number of mutual connections between client and server, which started up to 5 minutes after the end of an analyzed connection. & & X & X & & 62.04 \% \\
			
			ModTCPHdrLen   &     & $\bullet$ Modus of TCP header lengths in all traffic.  & & X & & & 0.00 \% \\
			
			UrgCntIn &    & $\bullet$ The number of TCP URG flags occurred in inbound traffic. & & X & & & 0.00 \% \\
			
			FinCntIn &    & $\bullet$ The number of TCP FIN flags occurred in inbound traffic. & & & X & & 53.76 \% \\
			
			PshCntIn &    & $\bullet$ The number of TCP PUSH flags occurred in inbound traffic. & & & X & & 54.00 \% \\
			
			FourGonModulIn[1]   &    & $\bullet$ Fast Fourier Transformation (FFT) of inbound packet sizes. The feature represents the module of the 2nd coefficient of the FFT in goniometric representation. & & X & X & & 95.15 \% \\
			
			FourGonModulOut[1]   &    & $\bullet$ The same as the previous one, but it represents the module of the 2nd coefficient of the FFT for outbound traffic. & & & X & & 99.50 \% \\
			
			FourGonAngleOut[1]   &    & $\bullet$ The same as the previous one, but it represents the angle of the 2nd coefficient of the FFT. & & X & & & 99.51 \% \\
			
			FourGonAngleN[9]   &    & $\bullet$ Fast Fourier Transformation (FFT) of all packet sizes, where inbound and outbound packets are represented by negative and positive values, respectively. The feature represents the angle of the 10th coefficient of the FFT in goniometric representation. & & X & X & & 99.69 \% \\	
			
			FourGonAngleN[1]   &    & $\bullet$ The same as the previous one, but it represents the angle of the 2nd coefficient of the FFT. & & X & & & 99.69\% \\
			
			FourGonModulN[0]   &    & $\bullet$ The same as the previous one, but it represents the module of the 1st coefficient of the FFT. & & & X & & 98.80 \%  \\
			
			PolyInd13ordOut[13]   &    & $\bullet$ Approximation of outbound communication by polynomial of 13th order in the index domain of packet occurrences. The feature represents the 14th coefficient of the approximation.  & & X & & & 66.21 \%  \\
			
			PolyInd3ordOut[3]   &    & $\bullet$ The same as the previous one, but it represents the 4th coefficient of the approximation.  & & & X & & 99.50 \% \\
			
			GaussProds8All[1] &    & $\bullet$ Normalized products of all packet sizes with 8 Gaussian curves. The feature represents a product of the 2nd slice of packets with a Gaussian function that fits to the interval of the packets' slice.   & & X & & & 95.54 \% \\				
			
			GaussProds8Out[7] &    & $\bullet$ The same as the previous one, but computed above outbound packets and represents a product of the 8th slice of packets with a Gaussian function that fits to the interval of the packets' slice. & & & X & & 52.87 \% \\  
			
			InPktLen1s10i[5]   &    & $\bullet$ Lengths of inbound packets occurred in the first second of a connection, which are distributed into 10 intervals. The feature represents totaled outbound packet lengths of the 6th interval. & & X & & & 14.69 \% \\
			
			OutPktLen32s10i[3] &    & $\bullet$ The same as the previous one, but computed above the first 32 seconds of a connection.  The feature represents totaled outbound packet lengths of the 4th interval.  & &  & X & & 38.20 \% \\
			
			OutPktLen4s10i[2] &    & $\bullet$ The same as the previous one, but computed above the first 4 seconds of a connection.  The feature represents totaled outbound packet lengths of the 3rd interval.  & &  & X & & 35.83 \% \\
			
			\midrule
			\multicolumn{3}{l}{Average of Divergent Obfuscated Attacks (FFS DL)}  & & & & & 72.80 \%	\\
			\multicolumn{3}{l}{Average of Divergent Obfuscated Attacks (FFS DOL)} & & & & & 66.33 \%	\\					
			\bottomrule    
			
		\end{tabular}
		}
		\smallskip
		\caption{TCP connection-level features selected by FFS (Na\"ive Bayes with kernels)}
		\label{FFS-features}
\end{table*}

\end{appendices}

\end{document}